\documentclass[10pt,pre,aps,floats,superscriptaddress,floatfix,twocolumn,longbibliography]{revtex4-2}
\newcommand{\cmmnt}[1]{}
\usepackage[T1]{fontenc}
\usepackage{mlmodern}
\usepackage{amssymb,amsmath}
\usepackage{braket}
\usepackage{graphicx}
\usepackage{dcolumn}
\usepackage{subfigure}
\usepackage{tabularx}
\usepackage{appendix}
\usepackage[english]{babel}
\usepackage{bm}
\usepackage{comment}
\usepackage{makecell}
\usepackage{changepage}
\usepackage{multirow}
\usepackage[table]{xcolor}
\usepackage[mathscr]{euscript}
\usepackage{csquotes}
\usepackage{soul} % Removed the redundant second xcolor call here

\usepackage[colorlinks=true,linkcolor=blue,allcolors=blue]{hyperref}
\usepackage[noabbrev]{cleveref}

\begin{document}
\title{Phase separation, morphology, and metastability in the three-dimensional active Potts model}

\author{Aditya Kumar Dutta}
\email{saisakd2137@iacs.res.in}
\affiliation{School of Physical Sciences, Indian Association for the Cultivation of Science, Kolkata -- 700032, India.}

\author{Raja Paul}
\email{ssprp@iacs.res.in}
\affiliation{School of Physical Sciences, Indian Association for the Cultivation of Science, Kolkata -- 700032, India.}

\begin{abstract}
We study the three-dimensional active Potts model (APM) on a cubic lattice with 2, 4, and 6 orientational states. The steady-state phase diagrams exhibit low-density gaseous, high-density polar liquid, and liquid--gas coexistence phases, with the phase behavior controlled by density, temperature, and propulsion speed and retaining the main features of the previously studied two-dimensional APM. At weak propulsion, the dense phase forms a slab perpendicular to its direction of motion. With increasing propulsion speed, this morphology persists for the 2-state model, whereas the dense phases for 4 and 6 states reorient and propagate along their longitudinal direction. For the 6-state model, the reoriented dense phase further develops a cylindrical morphology. We also examine the stability of polar liquids against artificially introduced droplets using microscopic simulations and coarse-grained hydrodynamic equations. At high propulsion speeds, counter-propagating droplets can partially modify the initial polar state, while transverse droplets can completely alter it, demonstrating the metastable nature of polar liquids in the three-dimensional active Potts model.
\end{abstract}

\maketitle
\section{Introduction}\label{sec:intro}

Active matter physics studies the collective movement of large assemblies of individual, energy-consuming particles. These agents self-propel and exhibit synchronized motion, spontaneously organizing into coherent structures on length scales significantly larger than the individual constituents. This clustering of active particles, known as flocks, is driven by an out-of-equilibrium phenomenon known as the flocking transition~\cite{Rama, Shae, Magis, Menon}. Such behavior is observed extensively in both the biological~\cite{March,Bot, Hel,Garci,Ball,Becco,Calovi,Steager,Peru,Giavazzi,Schaller,Sumino,Sanchez} and synthetic~\cite{DeseigneCombo,Deseigne3,Bric} worlds.

In the context of studying flocking transition, to circumvent the analytical and computational challenges posed by the prototypical model introduced by Vicsek et al.~\cite{Vicsek}, the active Ising model (AIM) was developed~\cite{solon2015flocking,AIM_solon_intro}, replacing the continuous symmetry of the Vicsek model (VM) by discrete rotational symmetry. Akin to a dynamical Ising model, it has two main ingredients for flocking of spins within a two-dimensional square lattice with coordination number 4 -- local alignment interactions and self-propulsion via biased hopping to neighboring sites without repulsive interactions. It has simpler, more tractable behavior and retains a large part of the physics in the VM, for instance, the two homogeneous states -- a disordered \enquote{gas} and an ordered \enquote{liquid}, the latter experiencing a collective migration of particles towards the two preferred directions (left or right) for self-propulsion. The flocking transition between these two states was interpreted as a phase-separation transition (rather than a simple order-disorder transition) featuring an intermediate coexistence phase where a liquid macroscopic band travels through a disordered gaseous background. 

A generalized active spin version of the AIM, the $q$-state active Potts model (APM), was addressed by Chatterjee et al.~\cite{chatterjee2020APM,SwarnajitAPM}. For $q>2$ in a square lattice, the flipping and hopping rules of the AIM are adjusted to accommodate the symmetry and physical properties of the Potts framework. In the coexistence phase, the band can move either transversely or longitudinally relative to its internal polarization. While the zero self-propulsion limit of the AIM belongs to the standard Ising universality class, the zero-propulsion limit of the 4-state APM does not fall into the universality class of the equilibrium 4-state Potts model.

Natural active systems predominantly operate in unconfined three-dimensional spaces. Simulations in three dimensions are often required to accurately model isotropic spatial sensing and out-of-plane degrees of freedom, which significantly alter collective evasion and spatial formations compared to their planar projections~\cite{cavagna2017dynamic,attanasi2014finite,ramlall2025role}. For instance, in continuous three-dimensional Vicsek models, restricting a particle's view optimizes directional consensus~\cite{li20113DVM}, while varying this vision range alongside noise induces structural transitions among gas, crystal, liquid-sphere, and vortex patterns~\cite{wu20213DVM}. Extending discrete flocking models into three dimensions allows the development of significantly more complex morphological patterns. Here, we ask how much of the AIM and APM phase behavior is qualitatively retained in their three-dimensional counterparts.

Beyond mapping the basic phase diagrams, a critical aspect of these discrete flocking models is the dynamical stability of their ordered phases. Active systems like the VM, AIM, or APM are capable of exhibiting long-range order (LRO) and can withstand spin-wave fluctuations~\cite{gregoire2004onset,bertin2006boltzmann,chate2008VM_variations,chate2008collective,bertin2009hydrodynamic} due to being intrinsically driven out-of-equilibrium. This led to the assumption that polar ordered phases in active matter are typically stable against large spontaneous density and orientational fluctuations~\cite{Toner1995LROXY,toner2005170,toner2012reanalysis,chate2008collective,mahault2019quantitative}. However, recent findings have shown that even a small obstacle can destabilize a flock, triggering transient dynamics before the system reaches a new steady state~\cite{Codina2022obstacle,Benvegnen2023metastability,chatterjee2025APMmetastability}. Similarly, rare but large spontaneous fluctuations~\cite{Besse2022metastability,woo2024MIP,mangeat2025TSAIM,dopierala2025NRXY,chatterjee2025APMmetastability} have also been observed to disrupt the LRO phase. 

In contrast to the continuous VM, which remains stable below a critical noise level~\cite{Codina2022obstacle}, the ordered phase of the discrete-symmetry AIM exhibits pronounced metastability -- both when subjected to artificially nucleated minority-phase droplets~\cite{Benvegnen2023metastability} and through the spontaneous formation of droplets with opposite polarity~\cite{woo2024MIP,mangeat2025TSAIM}. This suggests that the entire flocking phase of the AIM is strictly metastable in the thermodynamic limit~\cite{Benvegnen2023metastability}. This fragility extends to the 4-state APM~\cite{chatterjee2020APM,SwarnajitAPM}; despite possessing more orientational degrees of freedom, the polar liquid phase is readily perturbed by small counterpropagating or transversely propagating droplets~\cite{chatterjee2025APMmetastability}. Specifically, under conditions of low diffusion and high directional bias, a globally ordered flock can fragment into multiple smaller domains corresponding to the four internal spin states due to the spontaneous nucleation of these transverse droplets~\cite{chatterjee2025APMmetastability}.

Recent hydrodynamic theories predict that $d > 2$ AIM systems, despite exhibiting prolonged transient lifetimes without significant changes in initial droplet size, are ultimately not immune to metastability either~\cite{Benvegnen2023metastability}. In this paper, we phenomenologically explore how this vulnerability scales in the $d=3$ four-state APM. By artificially inserting a competing droplet into a fully ordered liquid background, we utilize time-evolution snapshots to track its spatial dynamics and visually assess whether the fragmentation mechanisms observed in two dimensions persist in three spatial dimensions.

This paper is organized as follows. In Sec.~\ref{sec:model}, we discuss the models and then present the details of numerical simulations in Sec.~\ref{sec:sim_details}. In Sec.~\ref{sec:num_results}, we will briefly study the phase space of three-dimensional active spin systems in cubic lattices, namely: (a) 2-state active Potts model ($q=2$, 3d-APM), which is equivalent to the active Ising model (3d-AIM), (b) 4-state active Potts model ($q=4$, 3d-APM) and (c) 6-state active Potts model ($q=6$, 3d-APM), followed by a visual metastability analysis of (b). Finally, in Sec.~\ref{sec:discuss}, we conclude this paper with a summary and discussion of the results.

\begin{figure*}[t]
    \centering
    \includegraphics[width=\textwidth]{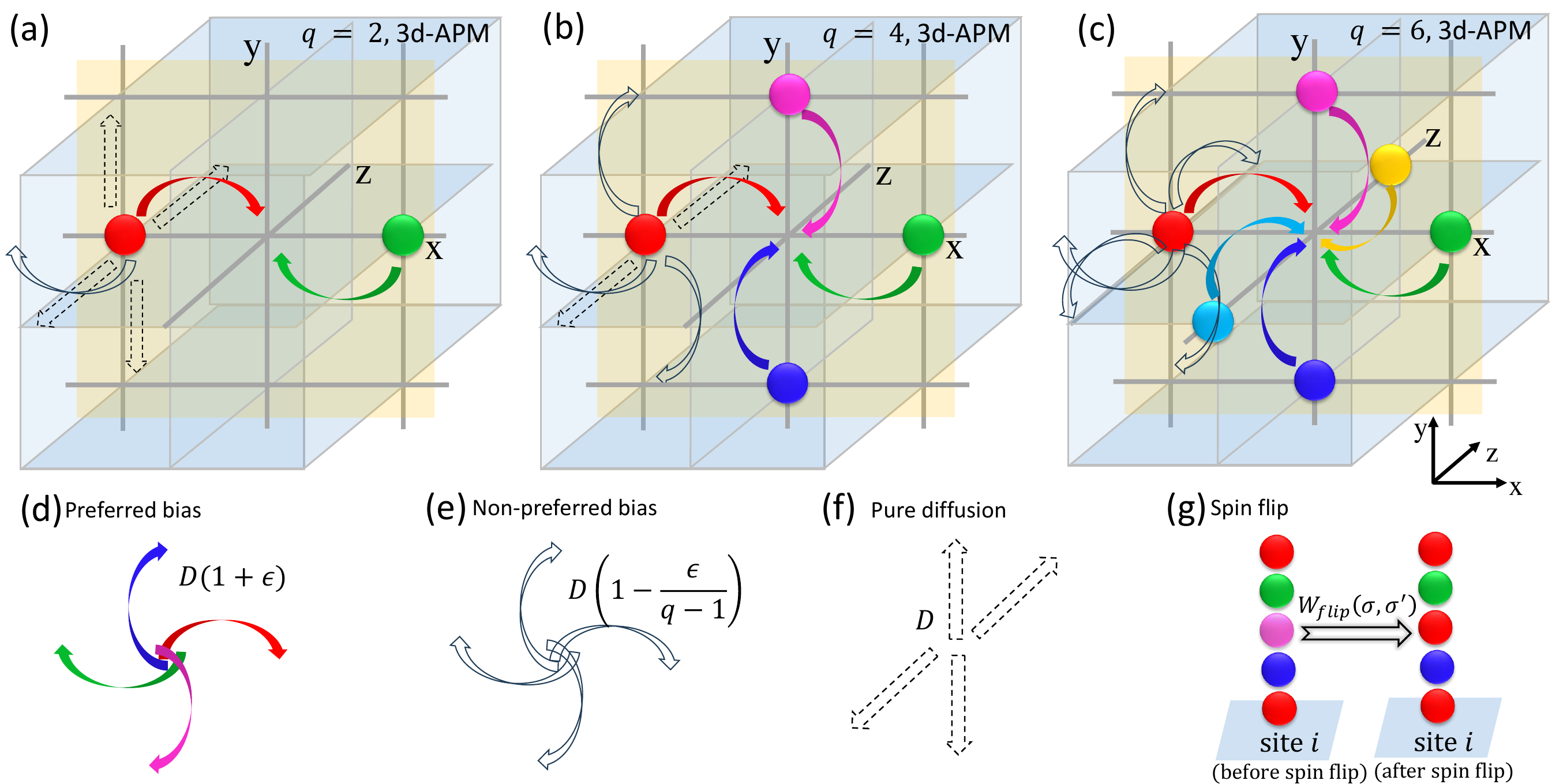}
   \caption{\textbf{Schematics of 3d-active Potts model dynamics.} (a--c) Particles hop between nearest-neighbor lattice sites in the (a) 2-state, (b) 4-state and (c) 6-state 3d active Potts model. Several particles can occupy the same site. Filled curved arrows denote preferred hopping, while open arrows indicate non-preferred hopping. For the \(q=2\) and \(q=4\) models, active hopping occurs in the \(xy\) plane; the \(q=6\) model also has active hopping along \(z\). Dashed arrows represent pure diffusion. For clarity, hopping and diffusion are illustrated only for the red particle. (d--f) Schematic of the corresponding hopping and diffusion rates. (g) Local spin reversal from state $\sigma$ to state $\sigma^\prime$ occurs with rate $W_{\rm flip}(\sigma,\sigma^\prime)$.}
    \label{fig:figlat}
\end{figure*}

\section{Microscopic models}
\label{sec:model}

Let us consider an ensemble of $N$ particles defined on a cubic lattice of dimensions $L_x \times L_y \times L_z$ with coordination number $Z= 6$. Each particle is in one of $q$ discrete internal states corresponding to a movement in one of the six lattice directions. Multiple particles can occupy each site, and each particle can either flip to a different spin state or hop to the nearest neighbor site. The alignment (i.e., flip) probabilities for particles on site $i$ are defined by the local Hamiltonian:
\begin{equation}
H=-\frac{J}{2\rho_i}\sum_{k=1}^{\rho_i}\sum_{l\neq{k}}(q\delta_{\sigma_{k},\sigma_{l}} - 1) 
\label{eq:hamiltonian}
\end{equation}
where the double sum runs over all particle pairs ($k,\ l$) on site $i$, $\sigma_k$ denotes the state of the $k$-th particle on site $i$: $\sigma_k \in \{1,\ \dots,\ q\}$, and $J$ is the coupling strength constant between different particles on the same site. The number of particles on site $i$ is $\rho_i = \sum_{\sigma=1}^{q}n_i^\sigma$ with $n_i^\sigma = \sum_{k=1}^{\rho_i}\delta_{\sigma_{k},\sigma}$, the number of particles in state $\sigma$. The local magnetization in the direction $\sigma$ on site $i$ is:
\begin{equation}
m_i^\sigma=\sum_{k=1}^{\rho_i}\frac{(q\delta_{\sigma,\sigma_{k}} - 1)}{q - 1}=\frac{qn_i^\sigma - \rho_i}{q - 1}\, .
\label{eq:magnetization}
\end{equation}
From Eq.~\eqref{eq:hamiltonian} and Eq.~\eqref{eq:magnetization}, we retrieve the local Hamiltonian and magnetization expressions for the AIM~\cite{solon2015flocking} ($q=2$) and APM~\cite{SwarnajitAPM} ($q=4$). A particle on site $i$ in state $\sigma$ changes its state to $\sigma^\prime$ with the rate $W_{\rm flip}(\sigma,\ \sigma^\prime) \propto \exp(-\beta \Delta H_{i})$, where $\beta = 1/T$ and $\Delta H_i = qJ(n_i^\sigma - n_i^{\sigma^\prime} - 1)/\rho_i$. Moreover, the particle performs a biased diffusion and jumps to the neighboring site in the direction $p \in \{1,\ \dots,\ 6\}$ with the rate $W_{\rm hop}(\sigma,\ p) \propto D[1 + h\epsilon(q\delta_{\sigma,p} - 1)/(q - 1)]$ where $\epsilon$ is the self-propulsion parameter. Meanwhile, $h$ is the hopping parameter, which is equal to 1 for the $q$ directions where biased diffusion takes place and is equal to 0 for the pure diffusion in the remaining $6 - q$ directions.

\subsection*{Dynamical rules for 3d-APM}
\label{sec:3DAIMdynam}
Here, we adopt the model introduced in~\cite{chatterjee2020APM,SwarnajitAPM} and obtain the on-site spin-flipping rates from the Potts Hamiltonian [Eq.~\eqref{eq:hamiltonian}]. Depending on their internal state (spin) and velocity, particles have a preferred direction for hopping to neighboring lattice sites. The hopping rate is higher in the preferred direction than in the non-preferred directions. In addition, particles undergo diffusive hopping along the remaining directions, which is independent of their internal state and velocity. Thus, by construction, in the $q=2$, 3d-APM, particles preferentially hop along the $x$-axis, while diffusion occurs along the $y$- and $z$-axes. For $q=4$, particles preferentially hop within the $xy$-plane and diffuse along the $z$-axis. In the $q=6$ model, particles can preferentially hop along all three spatial axes. Particles at each lattice site can additionally undergo spin flips with a specified rate. The hopping and spin-flipping processes are schematically illustrated in Fig.~\ref{fig:figlat}(a--g). Below, we provide the hopping and spin-flipping rates for a general $q$-state 3d-APM.

\begin{subequations}
\begin{align}
&W_{\rm flip}(\sigma, \sigma^\prime) &= \exp\left[- \frac{q \beta J}{\rho_i}(n_i^\sigma - n_i^{\sigma^\prime} - 1) \right] \label{eq:subeq5}\\
&W_{\rm hop}^\text{preferred bias} &= D(1+\epsilon) \label{eq:subeq6}\\
&W_{\rm hop}^\text{non-preferred bias} &= D\left(1-\frac{\epsilon}{q-1}\right) \label{eq:subeq7}\\
&W_{\rm hop}^\text{purely diffusive} &= D \label{eq:subeq8}
\end{align}
\end{subequations}

\section{Simulation details}
\label{sec:sim_details}

A Monte Carlo (MC) simulation of the stochastic process defined in this way evolves in unit Monte Carlo steps (MCS) $\Delta t$ resulting from a microscopic time $\Delta t/N$. During $\Delta t/N$, a randomly chosen particle either updates its spin state with probability $p_{\rm flip} = W_{\rm flip}(\sigma,\ \sigma^\prime)\Delta t$ or hops to one of the 6 neighboring sites with probability $p_{\rm hop} =\sum_{p=1}^{6} W_{\rm hop}(\sigma,\ p)\Delta t = 6D\Delta t$. The probability that nothing happens during this time is represented by $p_{\rm wait} = 1 - p_{\rm flip} - p_{\rm hop}$. An expression for $\Delta t$ can be obtained by minimizing $p_{\rm wait}$:
\begin{equation}
\begin{aligned}[b]
p_{\rm flip} + p_{\rm hop} &= \left \{6D + \gamma \exp\left[- \frac{q \beta J}{\rho_i}(n_i^\sigma - n_i^{\sigma^\prime} - 1) \right] \right \}\Delta t \\
&< \left[6D + \gamma \exp(q \beta J) \right]\Delta t
\end{aligned}
\label{eq:totalprob}
\end{equation}
Therefore, 
\begin{equation}
\Delta t = \left[6D + \gamma \exp(q \beta J) \right]^{-1}
\label{eq:4}
\end{equation}
Unless otherwise specified, we set $\gamma$, $D$, and $J$ to 1 for all models.

\section{Numerical Results}
\label{sec:num_results}

We mainly inspect the coexistence regime, where the system phase separates into a band of polar liquid (density $\sim \rho_{l}$) that travels in one of two directions through a disordered gaseous background (density $\sim \rho_{g}$). We plot the $T-\rho$ and $\epsilon-\rho$ phase diagrams navigating through the {$\beta,\epsilon,\rho$} parameter space for the models, where $\rho=N/L_{x}L_{y}L_{z}$ is the average density.

\subsection{Steady state features of \texorpdfstring{$q=2$}{q=2}, 3d-APM}
\label{sec:3DAIM}

\begin{figure*}[t]  
\centering
\includegraphics[width=\textwidth]{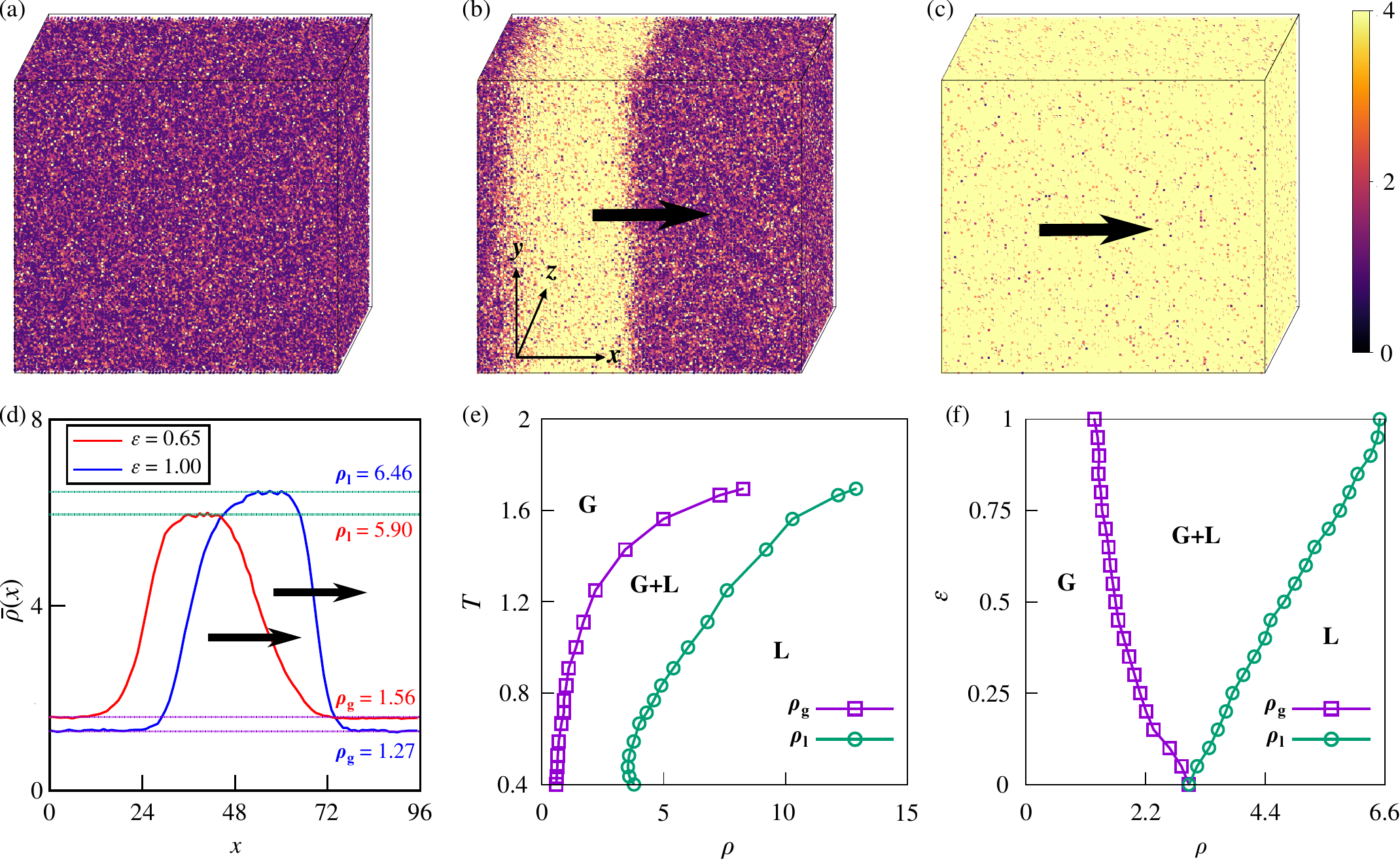}
    \caption{\textbf{2-state 3d-APM:} (a--c) Snapshots in $96 \times 96 \times 96$ domains with the colorbar representing site density, showing (a) gas ($\rho_0=1$), (b) gas-liquid coexistence ($\rho_0=3$) and (c) liquid ($\rho_0=8$) phases for $\beta = 1$, $\epsilon=1$. (d) YZ-averaged density profiles $\bar{\rho} (x)$ of the coexistence regime for two self-propulsion speeds ($\beta=1$). (e) Temperature-density ($T-\rho$) phase diagram for $\epsilon = 0.9$, and (f) Velocity-density ($\epsilon-\rho$)  phase diagram for $\beta =1$.
    }
    \label{fig:3DPD}
\end{figure*}

The $q=2$, 3d-APM exhibits three distinct phases: a disordered gaseous phase, a polar liquid phase, and a liquid--gas coexistence phase [Fig.~\ref{fig:3DPD}(a--c)]. These phases are qualitatively similar to those observed in the 2d-AIM~\cite{AIM_solon_intro,solon2015flocking}. In the coexistence regime, the system separates into a high-density ordered phase and a low-density disordered background. The dense phase forms a planar slab extending across the $yz$-plane and propagates through the surrounding gas. Its direction of motion is set by the dominant orientation of the particles within the slab. In the coexistence state shown in Fig.~\ref{fig:3DPD}(b), most particles in the slab are oriented along the $x$-direction, resulting in propagation along the $x$-axis. The width of the slab decreases with increasing propulsion speed.

To determine the gas and liquid binodals, $\rho_{\rm g}$ and $\rho_{\rm l}$, respectively, we obtain the density profile of the liquid--gas coexistence state by averaging over the $yz$-plane:
\begin{equation}
\bar \rho(x) = \frac{1}{L_y L_z} \sum_{y=0}^{L_y-1} \sum_{z=0}^{L_z-1} n(x, y, z)\, ,
\end{equation}
where $n(x, y, z)$ denotes the number of particles at the discrete lattice site $(x,y,z)$. The resulting density profile allows the gas and liquid densities to be identified from the corresponding low- and high-density regions. The binodals obtained in this way are shown in Fig.~\ref{fig:3DPD}(d).

For a fixed $\epsilon=0.9$, we construct the phase diagram in the $(T,\rho)$ plane [Fig.~\ref{fig:3DPD}(e)]. The binodals $\rho_{\rm g}$ and $\rho_{\rm l}$ delimit the gaseous (G), gas--liquid coexistence (G+L), and liquid (L) phases. We also obtain the phase diagram in the $(\epsilon,\rho)$ plane at fixed temperature $\beta=1$ [Fig.~\ref{fig:3DPD}(f)]. In this case, the two binodal lines converge at $\epsilon=0$ at a transition density $\rho^* \sim 3$ ($\beta=1$, $\epsilon=0$). At this density, the coexistence region disappears, and the system changes directly from the gaseous to the liquid state.

Compared with the 2d case, the larger number of particles and increased neighborhood connectivity in the 3d system suppress finite-size fluctuations, resulting in smoother density profiles. Additionally, at lower temperatures, the liquid binodal $\rho_{\rm l}$ shifts slightly towards higher densities than the conventional steady-state expectation.

\subsection{Steady state features of \texorpdfstring{$q = 4$}{q=4} state 3dAPM}

\begin{figure*}[t]  
\centering
\includegraphics[width=\textwidth]{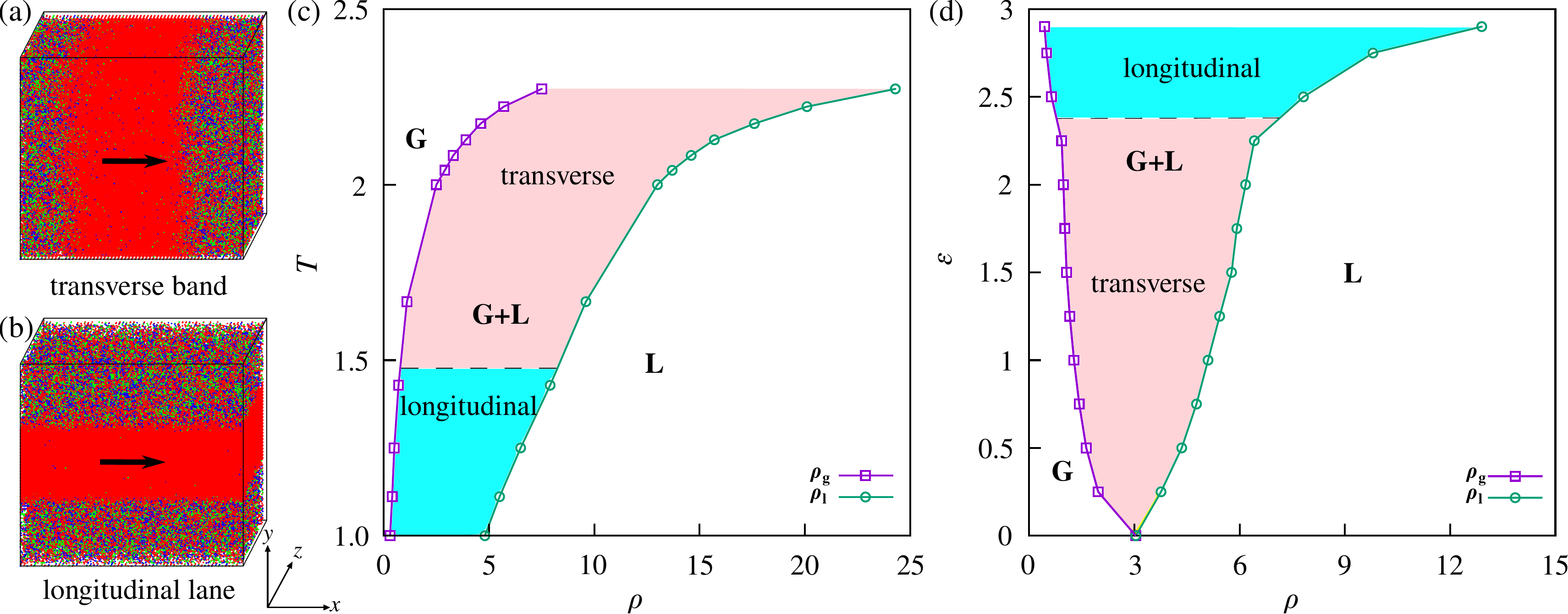}
    \caption{\textbf{4-state 3{d}-APM:} Orientational snapshots: (a) showing transverse band movement at low propulsion ($\epsilon=1$) and (b) showing longitudinal lane movement at high propulsion ($\epsilon=2.5$) for $\rho=3, \beta =0.7$ in a $96 \times 96 \times 96$ domain. Color mapping: right ($\sigma = 1$, red); up ($\sigma = 2$, blue); left ($\sigma = 3$, green); down ($\sigma = 4$, magenta). (c) Temperature-density ($T-\rho$) phase diagram for $\epsilon = 2.1$, and (d) Velocity-density ($\epsilon-\rho$) phase diagram for $\beta =0.7$. %\textcolor{red}{Correct the coordinate axes.}
    }
    \label{fig:3D4sPD}
\end{figure*}

Our numerical simulation for 3d active Potts model with $q=4$ shows that the system phase-separates into a condensed polar liquid traveling through a three-dimensional disordered gaseous background [Fig.~\ref{fig:3D4sPD}(a, b)]. This feature is similar to the two-dimensional scenario reported in~\cite{chatterjee2020APM,SwarnajitAPM}. However, in 3d, the liquid \enquote{band} manifests as a solid parallelepiped slab propagating along one of the four biased directions within the xy-plane. For lower biased propulsion $\epsilon$, the liquid forms a transverse slab (e.g., spanning the $yz$-plane while moving along the $x$-direction) [Fig.~\ref{fig:3D4sPD}(a)]. Conversely, for higher $\epsilon$, the system reorients into a longitudinal slab (e.g., spanning the xz-plane and moving along the $x$-direction) [Fig.~\ref{fig:3D4sPD}(b)].

Like in Sec.~\ref{sec:3DAIM}, the $T -\rho$ phase diagram is plotted for fixed $\epsilon=2.1$ [Fig.~\ref{fig:3D4sPD}(c)], with the binodals $\rho_{\rm g}$ and $\rho_{\rm l}$ delimiting the gaseous (G), gas-liquid coexistence (G+L), and liquid (L) phases. We find that the longitudinal lane (slab) at lower $T$ (higher $\beta$) switches to a transverse band (slab) at higher $T$ (lower $\beta$), with the reorientation transition happening at $T \sim 1.43$ ($\beta \sim 0.7$), represented by a horizontal black dotted line.

Meanwhile, in the $\epsilon -\rho$ phase diagram for $\beta=0.7$ [Fig.~\ref{fig:3D4sPD}(d)], this reorientation approximately happens at $\epsilon \geq 2.3$ (represented by a horizontal black dotted line), i.e., $\epsilon < 2.3$ is characterized by transverse band (slab) motion whereas $\epsilon > 2.3$ is characterized by longitudinal lane (slab) formation. At $\epsilon=0$, $\rho_{\rm g}$ and $\rho_{\rm l}$ merge at the transition density point $\rho^* \sim 3.1$ ($\beta=0.7$). The topological structure of these phase diagrams, including the existence and approximate location of the transverse-to-longitudinal reorientation transition, is consistent with the results obtained for the 2{d}APM~\cite{chatterjee2020APM,SwarnajitAPM}. This indicates that the fundamental thermodynamic phase behavior of the discrete active Potts model is preserved despite the addition of a purely diffusive third dimension.

\subsection{Steady state features of \texorpdfstring{$q = 6$}{q=6} state 3d-APM}

\begin{figure}[t]  
\centering
\includegraphics[width=\columnwidth]{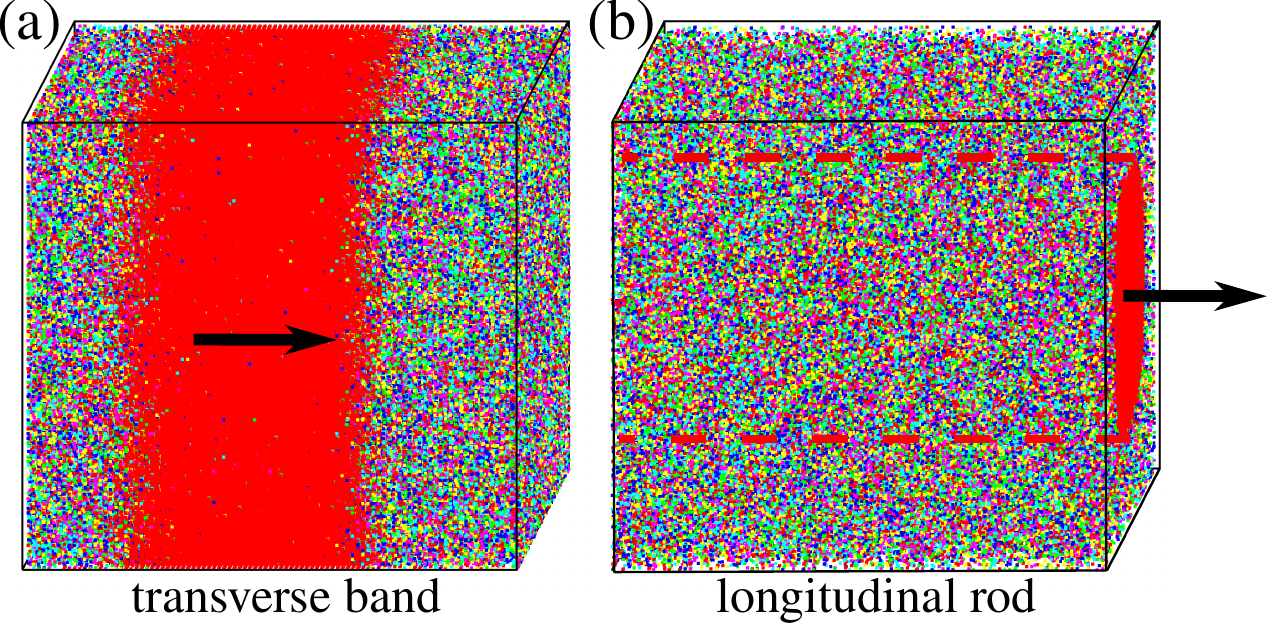}
    \caption{\textbf{6-state 3{d}-APM:} Orientational snapshots showing: (a) transverse band movement at low propulsion ($\epsilon=2.5$) and (b) longitudinal rod movement at high propulsion ($\epsilon=4.5$) for $\rho=2.5, \beta =0.6$ in a $64 \times 64 \times 64$ system. The dotted lines indicate the extent of the rod. Color mapping: right ($\sigma = 1$, red); up ($\sigma = 2$, blue); left ($\sigma = 3$, green); down ($\sigma = 4$, magenta); front ($\sigma = 5$, yellow); behind ($\sigma = 6$, cyan).}
    \label{fig:3D6sSNAP}
\end{figure}

Unlike the $q=2$ and $q=4$ models, the 6-state 3d-APM has no purely diffusive directions, since particles can actively hop along all three spatial axes. At low $\epsilon$, the system forms a moving transverse slab, similar to those observed for $q=2$ and $q=4$ [Fig.~\ref{fig:3D6sSNAP}(a)]. At higher propulsion, the slab reorients and develops into a longitudinal rod or cylindrical structure aligned with the direction of the majority state [Fig.~\ref{fig:3D6sSNAP}(b)], rather than the longitudinal slab observed for $q=4$ [Fig.~\ref{fig:3D4sPD}(b)]. This difference arises from the equal non-preferred hopping rate, $D(1-\epsilon/5)$, along all directions perpendicular to the direction of motion, which allows the dense phase to spread or shrink symmetrically around the propagation axis depending upon the propulsion parameter.

\begin{figure*}[t]  
\centering
\includegraphics[width=\textwidth]{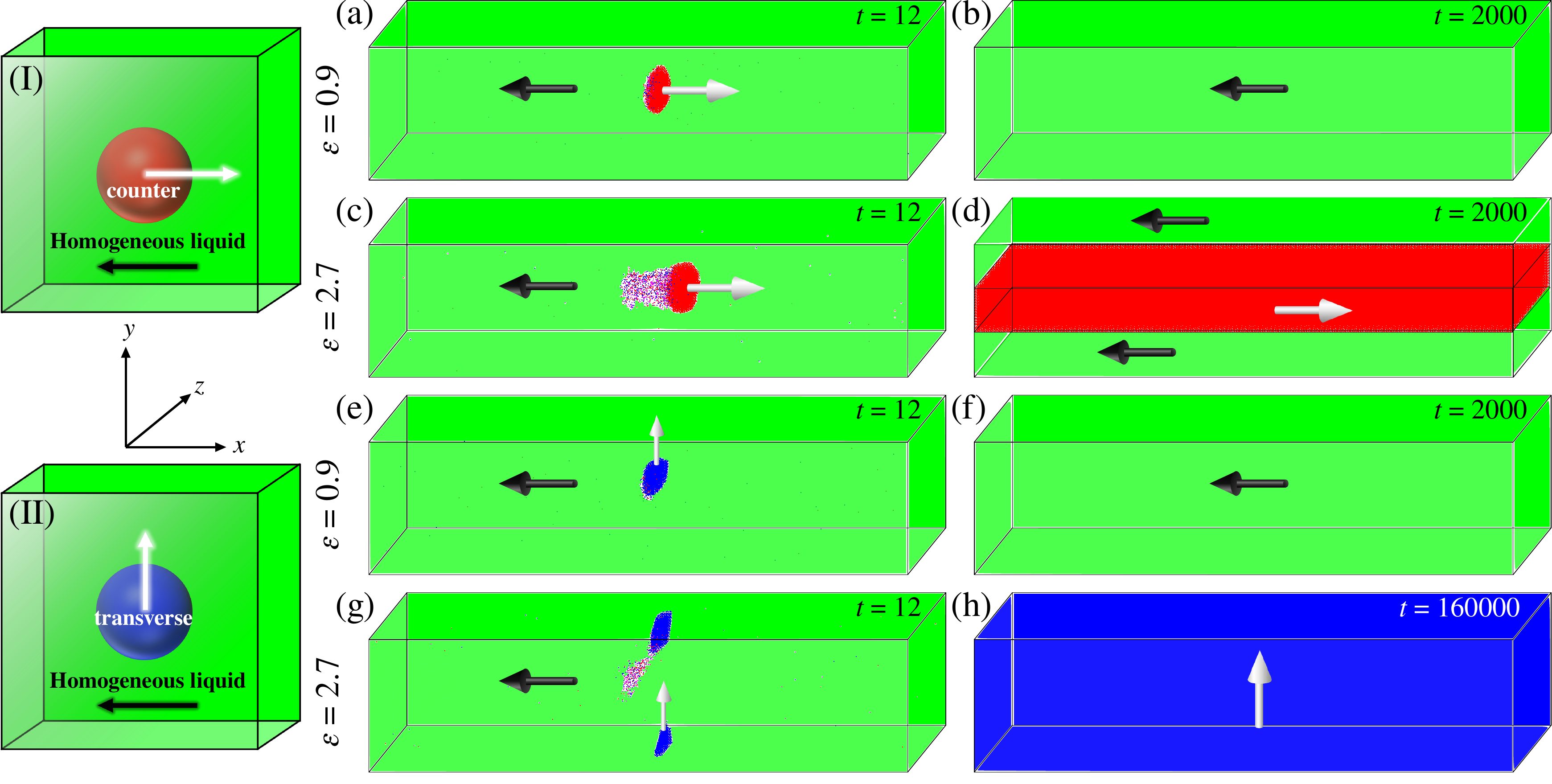}
    \caption{\textbf{Time evolution and steady states following droplet excitation in the 4-state 3d-APM} (I), (II) Schematics illustrating the insertion of (I) counterpropagating and (II) transversely propagating droplets into a homogeneous liquid background. Early-time (a, c, e, g) and steady-state (b, d, f, h) snapshots demonstrate the outcomes: (a, b) At low self-propulsion bias ($\epsilon=0.9$), a counterpropagating droplet dissolves into the background. (c, d) At high bias ($\epsilon=2.7$), it forms a persistent slab state comprising the droplet state (red) and background (green). (e, f) A transversely propagating droplet at low bias ($\epsilon=0.9$) similarly dissipates. (g, h) At high bias ($\epsilon=2.7$), the transverse droplet rapidly expands, altering the initial liquid phase. Parameters: $\beta = 1$, $\rho_0 = 10$, $r_d = 10$, $\rho_0^d = 1.2\rho_0$, $L_x = 500$, and $L_y = L_z = 50$. Color mapping: right ($\sigma = 1$, red); up ($\sigma = 2$, blue); left ($\sigma = 3$, green); down ($\sigma = 4$, magenta).}
    \label{fig:3D4smeta}
\end{figure*}

\subsection{Stability of the liquid phase for \texorpdfstring{$q = 4$}{q=4} state 3d-APM}
\label{sec:metastability}

We now examine the stability of the polar-order liquid phase by introducing either a counter-propagating [Fig.~\ref{fig:3D4smeta}(I)] or a transversely propagating [Fig.~\ref{fig:3D4smeta}(II)] liquid droplet into the high-density ordered liquid, following earlier simulation protocols~\cite{Benvegnen2023metastability,chatterjee2025APMmetastability}. We focus on $q=4$ because particles have significant degrees of freedom to hop to preferred and non-preferred directions and also to diffuse without bias. The initial homogeneous liquid is constructed entirely of $\sigma=3$ particles (green, propagating along $-x$) and thermalized by simulating over 150 time steps until it reaches the steady state. At this stage, the time is set to $t=0$ when a high-density spherical droplet of radius $r_d$, containing $\Delta N = 4/3(\rho_0^d -\rho_0) \pi r_d^3$ particles, is artificially injected at the center ($L_x/2, L_y/2, L_z/2$) of the simulation box. To test the response to competing polarities, the droplet's internal state is uniformly set to either $\sigma=1$ (counter-propagating against the background) or $\sigma=2$ (transversely propagating against the background). Keeping the inverse temperature ($\beta$), global density ($\rho_0$), and spatial dimensions constant, we evolve the system to a steady state ($t_{\rm eq} \sim 10^3 - 10^5$) across varying self-propulsion strengths ($\epsilon$).

Figures~\ref{fig:3D4smeta}(a--h) capture the time evolution and resulting spatial morphologies deep inside the liquid phase ($\beta=1, \rho_0 =10$). Notably, at a lower propulsion bias ($\epsilon=0.9$), a small counter-propagating droplet of state $\sigma=1$ (red, propagating along $+x$) [Fig.~\ref{fig:3D4smeta}(a, b)] fails to destabilize the polar background. This is a departure from the behavior of the 2d-APM~\cite{chatterjee2025APMmetastability}, where identical parameter settings readily disrupted the liquid state. This increased stability arises from the third spatial dimension ($L_z = 50$). The out-of-plane degree of freedom provides a purely diffusive pathway along the $z$-axis that increases non-longitudinal hopping, dispersing the droplet into the surrounding volume before a macroscopic phase reversal can nucleate. Conversely, when the self-propulsion bias is high ($\epsilon=2.7$), in-plane transverse hopping along the $y$-direction is heavily suppressed due to the vanishing $D(1-\epsilon/3)$ rate. The droplet is therefore unable to expand appreciably along the $y$-axis and instead extends predominantly along the direction of propagation. This results in a persistent, columnar slab configuration [Fig.~\ref{fig:3D4smeta}(c, d)], where the original ordered phase coexists alongside a counter-propagating planar slab that spreads across the entire $z$-dimension as a solid parallelepiped.

A similar dependence on spatial diffusion governs the transversely propagating droplets [Fig.~\ref{fig:3D4smeta}(e--h)]. Here, we inject a droplet of state $\sigma=2$ (blue, propagating along $+y$) into the $\sigma=3$ liquid background. At a low bias ($\epsilon=0.9$), out-of-plane diffusion along the purely diffusive $z$-axis (orthogonal to the active $xy$-plane) and in-plane transverse hopping dominate the droplet's boundaries, stripping away its internal order and dispersing it harmlessly into the stable background flow [Fig.~\ref{fig:3D4smeta}(e, f)]. However, at a high bias ($\epsilon=2.7$), the system's behavior changes: interfacial diffusion is minimized because the off-axis hopping rate $D(1-\epsilon/3)$ drops significantly. Consequently, the $\sigma=2$ droplet advances rapidly along its preferred $y$-axis. Its reduced interaction with the orthogonal background allows it to persist and expand across the $xy$-plane, eventually destabilizing the ordered state. This results in a total macroscopic phase reversal to a homogeneous $\sigma=2$ liquid [Fig.~\ref{fig:3D4smeta}(g, h)].

As explicitly addressed in the two-dimensional APM metastability study~\cite{chatterjee2025APMmetastability}, the complete structural reversal of a macroscopic domain by a counter-propagating droplet is inherently prevented by finite-size effects, leaving the sandwich configuration as the true steady-state outcome. Consistent with these observations, the inability of the counterpropagating droplet to induce full reversal in our three-dimensional environment is consistent with the greater stability observed in the three-dimensional system. Restricting the domain height to $L_z = 10$ [for the conditions in Fig.~\ref{fig:3D4smeta}(a, b)], or sufficiently increasing the droplet's size, density, and self-propulsion, consistently recovers the stable sandwich configuration.

\begin{figure}[t]  
\centering
\includegraphics[width=\columnwidth]{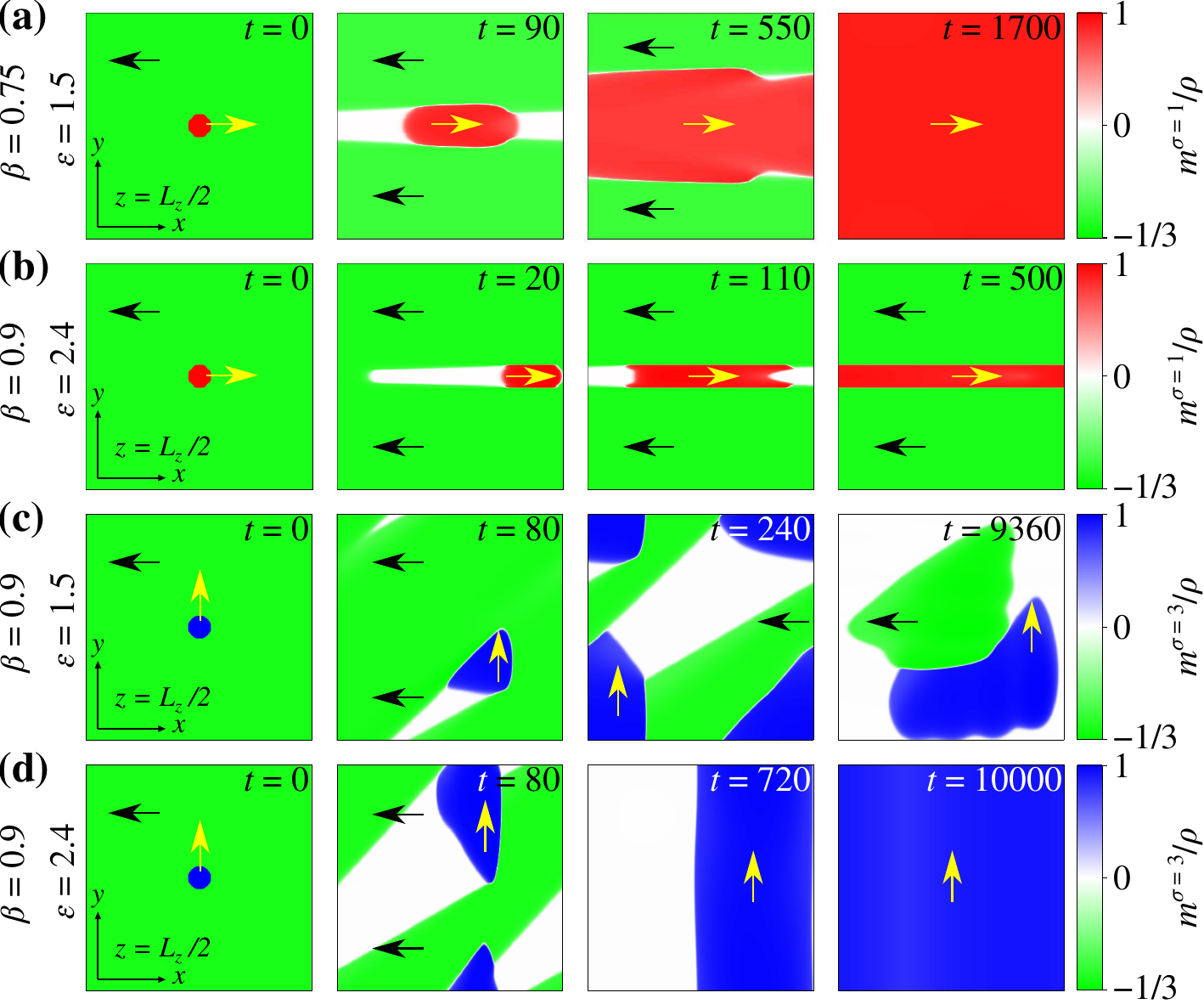}
    \caption{\textbf{Time evolution snapshots following hydrodynamic theory of the 4-state 3d-APM with droplet excitations, shown as cross-sectional slices in the $xy$ plane at $z=L_z/2$.} The initial ordered state is of $\sigma = 3$. (a, b) Counter-propagating ($\sigma = 1$) droplet leading to: (a) complete reversal ($\beta=0.75$, $\epsilon =1.5$) of initial ordered state and (b) formation of sandwich state ($\beta=0.9$, $\epsilon =2.4$). (c, d) Transversely propagating ($\sigma = 2$) droplet leading to: (c) formation of two persistent clusters with orthogonal states ($\beta=0.9$, $\epsilon =1.5$) and (d) complete reversal ($\beta=0.9$, $\epsilon =2.7$). Parameters: $D=1$, $\rho_0 = 3$, $r_d = 10$, $\rho_0^d = 5\rho_0$, and cubic system with $L = 200$. Arrows indicate the direction of motion, and the colorbar represents the corresponding droplet magnetization (Eq.~\eqref{eq:magnetization}).}
    \label{fig:3D4shydro}
\end{figure}

We now aim to discuss the results of droplet excitation for $q=4$ using a refined mean-field coarse-grained hydrodynamic description ~\cite{SwarnajitAPM}. Defining the average density field of particles in the state $\sigma$ at the position ${\bf r} =(x,y,z)$ as $\rho_\sigma ({\bf r},t) = \langle n_i^\sigma (t) \rangle$, its hydrodynamic equation is given by:
\begin{gather} 
    \partial_t\rho_\sigma = D_{\parallel} \partial^2_{\parallel} \rho_\sigma + D_{\perp} \partial^2_{\perp} \rho_\sigma + D \partial^2_z \rho_\sigma- v \partial_{\parallel} \rho_\sigma + \sum_{\sigma' \ne \sigma} I_{\sigma \sigma'}\, , 
    \label{rhoeq}   
\end{gather}
where $I_{\sigma \sigma'}$ is the flipping term defined as:
\begin{equation}
\begin{aligned}
I_{\sigma\sigma'} =
\Bigg[
    \frac{4\beta J}{\rho}
    \left(\rho_\sigma+\rho_{\sigma'}\right)
    -1-\frac{r}{\rho}
    -\alpha\frac{(\rho_\sigma-\rho_{\sigma'})^2}{\rho^2}
\Bigg]
\left(\rho_\sigma-\rho_{\sigma'}\right).
\label{flipterm}
\end{aligned}
\end{equation} 
$D_{\parallel} = D (1 +\epsilon/3)$ and $D_{\perp} = D (1 -\epsilon/3)$ are the diffusion constants in the $xy$-plane, along the directions ${\bf e}_{\parallel} = (\cos \phi,\sin \phi,0), {\bf e}_{\perp} = (\sin \phi, -\cos \phi,0)$ respectively, where $\phi = (\sigma -1)\pi/2$ is the favored angle for state $\sigma$. The self-propulsion velocity $v=4D \epsilon /3$ is along ${\bf e}_{\parallel}$. Derivatives in these directions are denoted as $\partial_\parallel = {\bf e}_\parallel \cdot \nabla$ and $\partial_\perp = {\bf e}_\perp \cdot \nabla$. In Eq.~\eqref{flipterm}, $\alpha = 8(\beta J)^2(1-2\beta J/3)$ and $r=27 \alpha_m \alpha/8$, where $\alpha_m = \alpha_m (\beta)$ is an unknown function of temperature. $\rho = \sum_{\sigma=1}^{4}\rho_\sigma$ is the total particle density. Formulation of Eq.~\ref{rhoeq} from the microscopic models is discussed in the Appendix~\ref{app:Hydrodynamic_eqn}.

Particles rely entirely on local majority rule (Eq.~\eqref{eq:subeq5}) to undergo spin flips. Because lattice sites contain a finite number of particles, these local majorities exhibit strong magnetization fluctuations. The parameter $\alpha_m$ quantifies the strength of localized magnetization fluctuations. Qualitatively, introducing the purely diffusive $z$-axis increases the available spatial pathways for particles. We expect this enhanced dispersion to accelerate local decorrelation, effectively reducing the amplitude of the magnetization fluctuations ($\alpha_m$) compared to the two-dimensional case.

We use an explicit forward-time central-space (FTCS) finite-difference scheme to numerically integrate Eq.~\eqref{rhoeq}. We solve the four coupled partial differential equations on a three-dimensional cubic domain of size $L \times L \times L$ with $L = 200$, applying periodic boundary conditions in all three directions. The spatial mesh size is set to $\Delta x = 0.5$ and the integration time step to $\Delta t = 0.05\Delta x^2 =0.0125$, with maximum simulation time $t_{\text{sim}} = 10^4$. 

To ensure stability of the explicit integration, the time step must satisfy the diffusive von Neumann criterion~\cite{press2007numerical,strikwerda2004finite} across all spatial axes:
\begin{equation}
\Delta t \le \frac{\Delta x^2}{2(D_\parallel + D_\perp + D)}.
\end{equation}
Because the active corrections to diffusion cancel identically along the principal axes ($D_\parallel + D_\perp + D = 3D$), this stability threshold is independent of the activity parameter $\epsilon$, imposing $\Delta t \le \Delta x^2 / (6D) \approx 0.0417$. Our choice of $\Delta t = 0.0125$ strictly satisfies this bound while remaining well within the advective stability limit for all investigated values of $\epsilon$. 

In our numerical implementation, we fix $D = J = r = 1$, defining the characteristic units of time, energy, and density. The system is initialized with a spherical high-density droplet of radius $R_d = 10$ and density $\rho_d = 15$, polarized differently from the surrounding homogeneous liquid background ($\rho_0 = 3.0$).

Comparing the hydrodynamic results [Fig.~\ref{fig:3D4shydro}] to our discrete simulations demonstrates the impact of finite-size fluctuations. For a counter-propagating droplet at lower bias, the theory predicts a complete phase reversal [Fig.~\ref{fig:3D4shydro}(a)] and at higher bias, a sandwiched slab state [Fig.~\ref{fig:3D4shydro}(b)]. However, as established in two dimensions~\cite{chatterjee2025APMmetastability}, such complete reversals are finite-size effects that occur when an expanding domain wraps around the periodic transverse boundaries. In our discrete 3d model, the combination of intrinsic demographic noise and unsuppressed out-of-plane diffusion rapidly disperses the droplet before this expansion can occur, causing it to simply dissipate [Fig.~\ref{fig:3D4smeta}(a, b)]. 

Furthermore, for a transversely propagating droplet, we obtain orthogonally propagating clusters at lower bias [Fig.~\ref{fig:3D4shydro}(c)] and a full phase reversal at higher bias [Fig.~\ref{fig:3D4shydro}(d)]. While the continuous equations predict that such clusters can stably coexist [Fig.~\ref{fig:3D4shydro}(c)], this macroscopic geometric balance is unstable in the discrete system. Intrinsic demographic noise acting across the extensive three-dimensional intersection volume can break this symmetry, driving the discrete system toward a complete phase reversal instead [Fig.~\ref{fig:3D4smeta}(g, h)].

\section{Discussion}
\label{sec:discuss}

\begin{figure}[t]  
\centering
\includegraphics[width=\columnwidth]{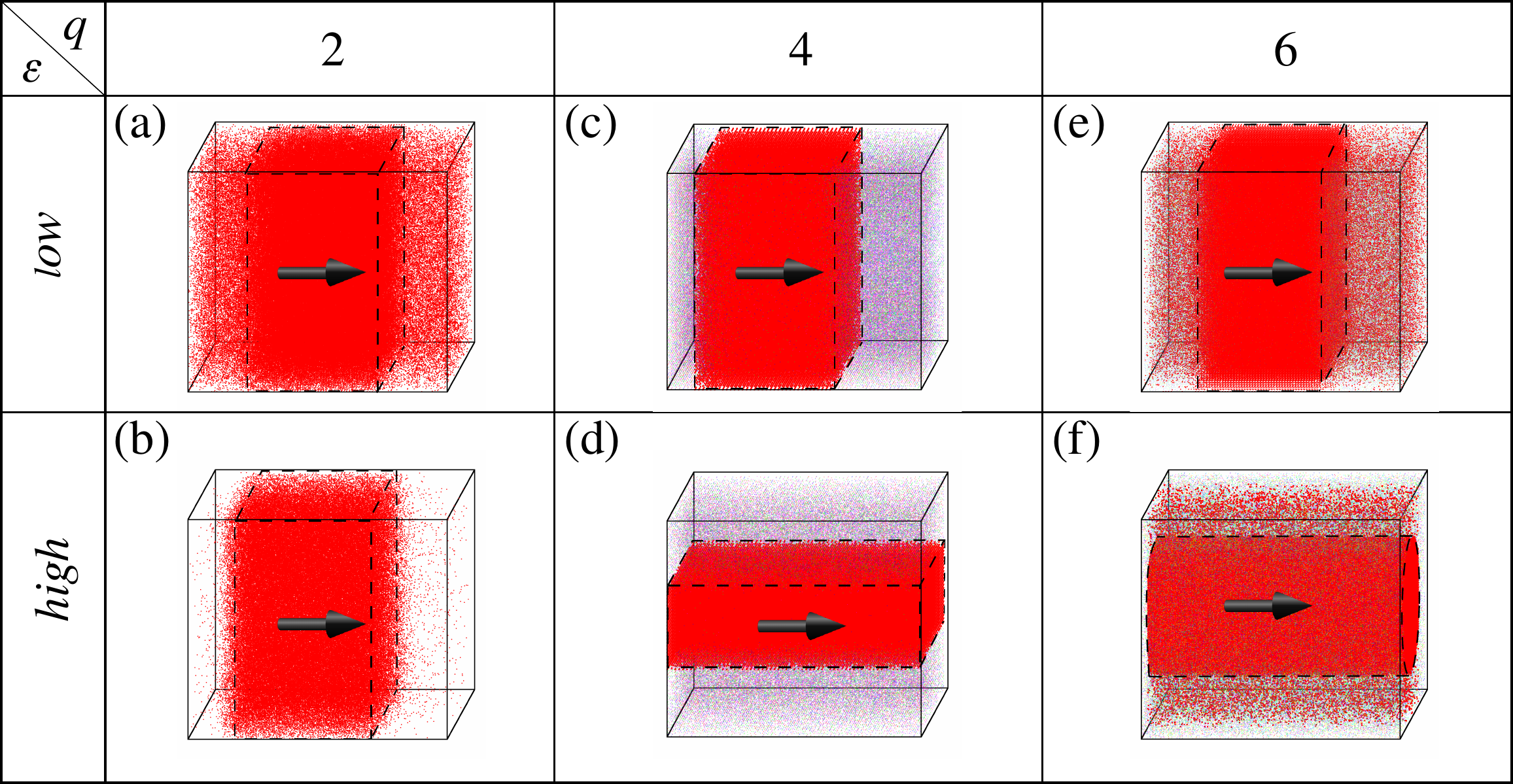}
    \caption{\textbf{Coexistence state morphology in 3d-APM:} (a, b) 2-state: dispersed band for $\epsilon=0.3$ in (a) and condensed band for $\epsilon=0.9$ in (b). (c, d) 4-state: transverse band for $\epsilon=1$ in (c) and longitudinal slab for $\epsilon=2.7$ in (d). (e, f) 6-state: transverse band for $\epsilon=2.5$ in (e) and longitudinal rod for $\epsilon=4.5$ in (f). System size $96\times 96 \times 96$.}
    \label{fig:3Dmorphosummary}
\end{figure}

In this article, we investigated the three-dimensional active Potts model (3d-APM) with $q=2$, $4$, and $6$ orientational states. We focused on two main aspects: (i) the collective states arising from the competition between self-propulsion and motion in non-preferred directions, and (ii) the response of these states to external perturbations. We studied these behaviors using microscopic simulations and corresponding refined mean-field coarse-grained hydrodynamic equations.

The 3d-APM retains the basic phase behavior observed in two dimensions, with disordered gaseous, ordered polar liquid, and polar-liquid--gas coexistence phases. In the coexistence regime, the morphology and orientation of the dense phase depend on both $q$ and the propulsion speed $\epsilon$. For $q=2$, the dense phase forms a slab perpendicular to its direction of propagation, and increasing $\epsilon$ mainly sharpens the slab profile while reducing its width [Fig.~\ref{fig:3Dmorphosummary}(a,b)]. For $q=4$, a transverse slab forms at low $\epsilon$ [Fig.~\ref{fig:3Dmorphosummary}(c)]. Here, active hopping is confined to the $xy$-plane, while motion along \(z\) is purely diffusive. At low $\epsilon$, the non-preferred active hopping and diffusion are comparable, forming a transverse slab. As $\epsilon$ increases, the non-preferred active hopping decreases while the diffusive motion remains unchanged, leading to an abrupt reorientation into a longitudinal slab [Fig.~\ref{fig:3Dmorphosummary}(d)].

For $q=6$, all six lattice directions allow active hopping, with no purely diffusive directions. At low $\epsilon$, relatively strong non-preferred hopping allows transverse spreading and produces a slab-like dense phase [Fig.~\ref{fig:3Dmorphosummary}(e)]. Increasing $\epsilon$ suppresses this transverse spreading, causing the slab to reorient and eventually develop into a longitudinal cylindrical or rod-like structure [Fig.~\ref{fig:3Dmorphosummary}(f)]. Thus, the number of orientational states determines the available directions of active and diffusive motion, while their relative mobility controls the morphology of the dense phase.

The additional spatial dimension also suppresses finite-size fluctuations by providing more directions for random hopping and particle redistribution. Consequently, density profiles are smoother, and the collective states are less sensitive to small perturbations than in two dimensions. However, this increased stability does not eliminate the metastability of polar liquids, consistent with recent hydrodynamic predictions for discrete-symmetry flocks~\cite{Benvegnen2023metastability}.

To examine metastability, we introduced high-density droplets composed of particles in an internal state different from that of the surrounding polar liquid. For $q=2$, a counter-propagating droplet disappears at low propulsion speed (see Appendix~\ref{metastability_3DAIM}), whereas a transverse droplet can substantially alter the original liquid state~\cite{Benvegnen2023metastability}. For $q=4$, a counter-propagating droplet similarly disappears at low propulsion speed, but at higher $\epsilon$ it can generate a reversed state that expands along the diffusive directions, forming a slab between regions of the original state. A transverse droplet can also alter the original state at high $\epsilon$. For $q=6$, a small counter-propagating droplet disappears at low $\epsilon$, whereas at higher $\epsilon$ it develops into a cylindrical liquid structure near the center of the simulation box. A transverse droplet can completely replace the original liquid state at high $\epsilon$ (see Appendix~\ref{metastability_3D6sAPM}), while at very high propulsion speeds a mixed state containing domains of several orientational states can form. The coarse-grained hydrodynamic equations reproduce the main features of the microscopic simulations, including the collective morphologies and their response to perturbations.

Taken together, our results show that the internal-state structure of the 3d-APM determines the available directions of active and diffusive motion, while their relative strengths control the morphology and stability of the collective states. In particular, counter-propagating perturbations at high propulsion can partially reverse the original polar state, producing a sandwich-like configuration, whereas transverse perturbations can lead to a more complete replacement of the original state.

Several questions remain open. The critical size and density of a perturbing droplet could provide a quantitative measure of polar-liquid stability, while the formation, growth, lifetime, and survival probability of spontaneous droplets could further characterize metastability. The three-dimensional model also provides a framework for studying active systems in intrinsically three-dimensional environments~\cite{attanasi2014finite,ramlall2025role}. Since low diffusion can drive spontaneous motility-induced pinning (MIP) through kinetic trapping in two dimensions~\cite{woo2024MIP,chatterjee2025APMmetastability}, an immediate extension would be to investigate this transition in three dimensions and determine whether it produces structurally richer jammed networks.

\section*{Acknowledgments}
A.D. sincerely acknowledges the Indian Association for the Cultivation of Science (IACS), Kolkata, India, for providing the fellowship and computational facilities. R.P. thanks IACS for its computational facilities and resources.
\appendix

\renewcommand{\appendixname}{APPENDIX}
\let\oldsection\section
\renewcommand{\section}[1]{\oldsection{\MakeUppercase{#1}}}

\section{Derivation of 4-state \MakeLowercase{3d}-APM hydrodynamic equation: hopping}
\label{app:Hydrodynamic_eqn}

From the dynamics of the 4-state 3d-APM, one can derive the master equation for the number of particles $n_i^\sigma (t)$ on site $i$ of state $\sigma$~\cite{chatterjee2020APM,SwarnajitAPM}:
\begin{widetext}
\begin{equation}
\langle n_i^\sigma (t+dt) \rangle = \Big\langle n_i^\sigma (t) \Big[ 1 - dt \sum_{p=1}^6 W_{\rm hop}(\sigma, p) - dt \sum_{\sigma' \ne \sigma} W_{\rm flip}(\sigma, \sigma') \Big] \Big\rangle + dt \Big\langle \sum_{p=1}^6 n_{i-p}^\sigma(t) W_{\rm hop}(\sigma, p) \Big\rangle + dt \Big\langle \sum_{\sigma' \ne \sigma} n_i^{\sigma'}(t) W_{\rm flip}(\sigma', \sigma) \Big\rangle
\label{app:master_eq}
\end{equation}
\end{widetext}

with subscript $i \pm p$ denoting the neighbor of site $i$ in the $\pm p$ direction. In the limit $dt \rightarrow 0$, the time evolution of $\langle n_i^\sigma \rangle$ due to spatial hopping is defined by the sum over all lattice directions:
\begin{equation}
\partial_t \langle n_i^\sigma \rangle \bigg|_{\text{hop}} = \sum_{p=1}^6 W_{\rm hop}(\sigma, p) [\langle n_{i-p}^\sigma \rangle - \langle n_i^\sigma \rangle]
\end{equation}

In the hydrodynamic limit for small lattice spacing $a \simeq 1/L$, we define the continuous density $\rho_\sigma(\bm{r}, t) = n_i^\sigma (t)$ at some coordinate $\bm{r} = (i_x,i_y,i_z)a$ and apply a Taylor expansion of neighboring density $\rho_\sigma(\bm{r} +a{\bm{\hat e}_p},t)$ up to the second order:
\begin{equation}
\begin{aligned}
n_{i+p}^\sigma &= \rho_\sigma(\bm{r} +a{\bm{\hat e}_p},t) \\
&= \rho_\sigma(\bm{r},t) + a \partial_p \rho_\sigma (\bm{r},t) + \frac{a^2}{2}\partial_p^2 \rho_\sigma (\bm{r},t)+ \mathcal{O}(a^3)
\end{aligned}
\end{equation}
where $\partial_p = \bm{\hat e}_p \cdot \nabla_{\bm r}$. The difference in particle number on neighboring sites is hence given by:
\begin{equation}
\langle n_{i-p}^\sigma \rangle - \langle n_i^\sigma \rangle \simeq -a \partial_p \rho_\sigma (\bm{r},t) + \frac{a^2}{2}\partial_p^2 \rho_\sigma (\bm{r},t) 
\end{equation}

To extend this to our three-dimensional configuration, we evaluate the spatial hopping flux over all six lattice directions simultaneously. For a state $\sigma$ with 4 active in-plane directions, the hopping rates are biased by the self-propulsion parameter $\epsilon$ within the active $xy$-plane, while the out-of-plane $z$-axis experiences unbiased symmetric diffusion. The directional hopping rates are defined as:
\begin{align}
W_{\rm F} &= D (1 + \epsilon) && \text{(forward, } \parallel \text{)} \\
W_{\rm B} &= D \left(1 - \frac{\epsilon}{3}\right) && \text{(backward, } -\parallel \text{)} \\
W_{\rm T} &= D \left(1 - \frac{\epsilon}{3}\right) && \text{(transverse, } \pm\perp \text{)} \\
W_{\rm Z} &= D && \text{(passive vertical, } \pm z \text{)}
\end{align}
Summing the contributions from these six directions and substituting the Taylor expansions gives the total spatial flux:
\begin{widetext}
\begin{equation}
\begin{aligned}
\text{Flux} &= W_{\rm F} \big( {-a} \partial_\parallel \rho_\sigma + \tfrac{a^2}{2} \partial_\parallel^2 \rho_\sigma \big) + W_{\rm B} \big( a \partial_\parallel \rho_\sigma + \tfrac{a^2}{2} \partial_\parallel^2 \rho_\sigma \big) + W_{\rm T} \big( {-a} \partial_\perp \rho_\sigma + \tfrac{a^2}{2} \partial_\perp^2 \rho_\sigma \big) \\
&\quad + W_{\rm T} \big( a \partial_\perp \rho_\sigma + \tfrac{a^2}{2} \partial_\perp^2 \rho_\sigma \big) + W_{\rm Z} \big( {-a} \partial_z \rho_\sigma + \tfrac{a^2}{2} \partial_z^2 \rho_\sigma \big) + W_{\rm Z} \big( a \partial_z \rho_\sigma + \tfrac{a^2}{2} \partial_z^2 \rho_\sigma \big)
\end{aligned}
\end{equation}
\end{widetext}
The first-order advective derivatives exactly cancel out for the symmetric transverse ($W_{\rm T}$) and vertical ($W_{\rm Z}$) directions. Grouping the remaining spatial derivatives, we obtain:
\begin{equation}
\begin{aligned}
\text{Flux} &= -a (W_{\rm F} - W_{\rm B}) \partial_\parallel \rho_\sigma \\
&\quad  + \frac{a^2}{2} (W_{\rm F} + W_{\rm B}) \partial_\parallel^2 \rho_\sigma \\
&\quad + a^2 W_{\rm T} \partial_\perp^2 \rho_\sigma + a^2 W_{\rm Z} \partial_z^2 \rho_\sigma
\end{aligned}
\end{equation}
Substituting the hopping rates into the flux equation, we get:
\begin{equation}
\begin{aligned}
\text{Flux} &= \frac{2D\epsilon}{3} a^2 \partial_\parallel^2 \rho_\sigma - \frac{4D\epsilon}{3} a \partial_\parallel \rho_\sigma \\ 
&\quad + D\left(1-\frac{\epsilon}{3}\right) a^2 \nabla_{xy}^2 \rho_\sigma  + D a^2 \partial_z^2 \rho_\sigma
\end{aligned}
\end{equation}
where we recognize the Laplacian $\nabla_{xy}^2 = \partial_\parallel^2 + \partial_\perp^2$ multiplied by the transverse active coefficient. Absorbing the lattice spacing $a$ into the macroscopic advection and diffusion constants ($v$, $D_\parallel$, $D_\perp$, and $D$), the spatial flux simplifies to:
\begin{equation}
\text{Flux} = D_\parallel \partial_\parallel^2 \rho_\sigma + D_\perp \partial_\perp^2 \rho_\sigma + D \partial_z^2 \rho_\sigma - v \partial_\parallel \rho_\sigma
\end{equation}

Finally, by equating the time derivative of the density to this spatial flux, we recover the hopping part of the hydrodynamic equation of the 4-state 3d-APM:
\begin{equation}
\partial_t\rho_\sigma\bigg|_{\text{hop}} = D_{\parallel} \partial^2_{\parallel} \rho_\sigma + D_{\perp} \partial^2_{\perp} \rho_\sigma + D \partial^2_z \rho_\sigma - v \partial_{\parallel} \rho_\sigma
\label{app:rhoeq}
\end{equation}

\section{Stability of the liquid phase for \texorpdfstring{$q = 2$}{q=2} state \MakeLowercase{3d}-APM}
\label{metastability_3DAIM}
\begin{figure}[t]  
\centering
\includegraphics[width=\columnwidth]{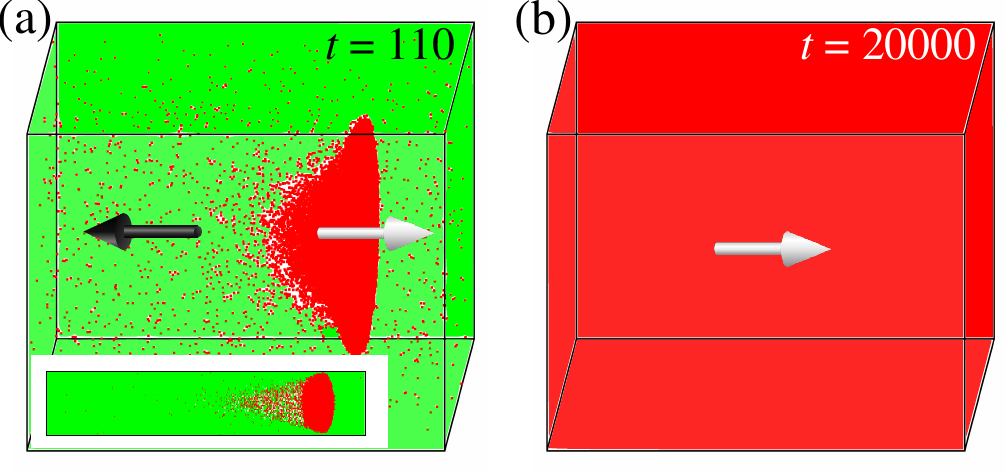}
    \caption{\textbf{Droplet excitation in the 2-state 3d-APM.} (a) Early time with inset showing $xy$-plane profile at $z=L_z/2$. (b) Full reversal at late time. Parameters: $\epsilon=0.9$, $\beta = 1$, $\rho_0 = 10$, $r_d = 10$, $\rho_0^d = 1.2\rho_0$, $L_x = 500$, and $L_y = L_z = 50$.}
    \label{fig:3D2smeta}
\end{figure}

Here we examine the stability of the polar-order liquid phase in the 2-state 3d-APM~\cite{Benvegnen2023metastability} by introducing a counter-propagating ($\sigma=1$) liquid droplet [Fig.~\ref{fig:3D4smeta}(I)] into the high-density ordered liquid ($\sigma=2$). 

Fig.~\ref{fig:3D2smeta} captures the time evolution at a high propulsion bias ($\epsilon=0.9$). The small counter-propagating droplet of state $\sigma=1$ (red, propagating along $+x$) expands into the surrounding ordered phase, leaving behind a dilute, disordered wake [Fig.~\ref{fig:3D2smeta}(a)], and eventually reverses the macroscopic order (green, propagating along $-x$) to match the droplet [Fig.~\ref{fig:3D2smeta}(b)].

\section{Stability of the liquid phase for \texorpdfstring{$q = 6$}{q=6} state \MakeLowercase{3d}-APM}
\label{metastability_3D6sAPM}

\begin{figure}[!t]  
\centering
\includegraphics[width=\columnwidth]{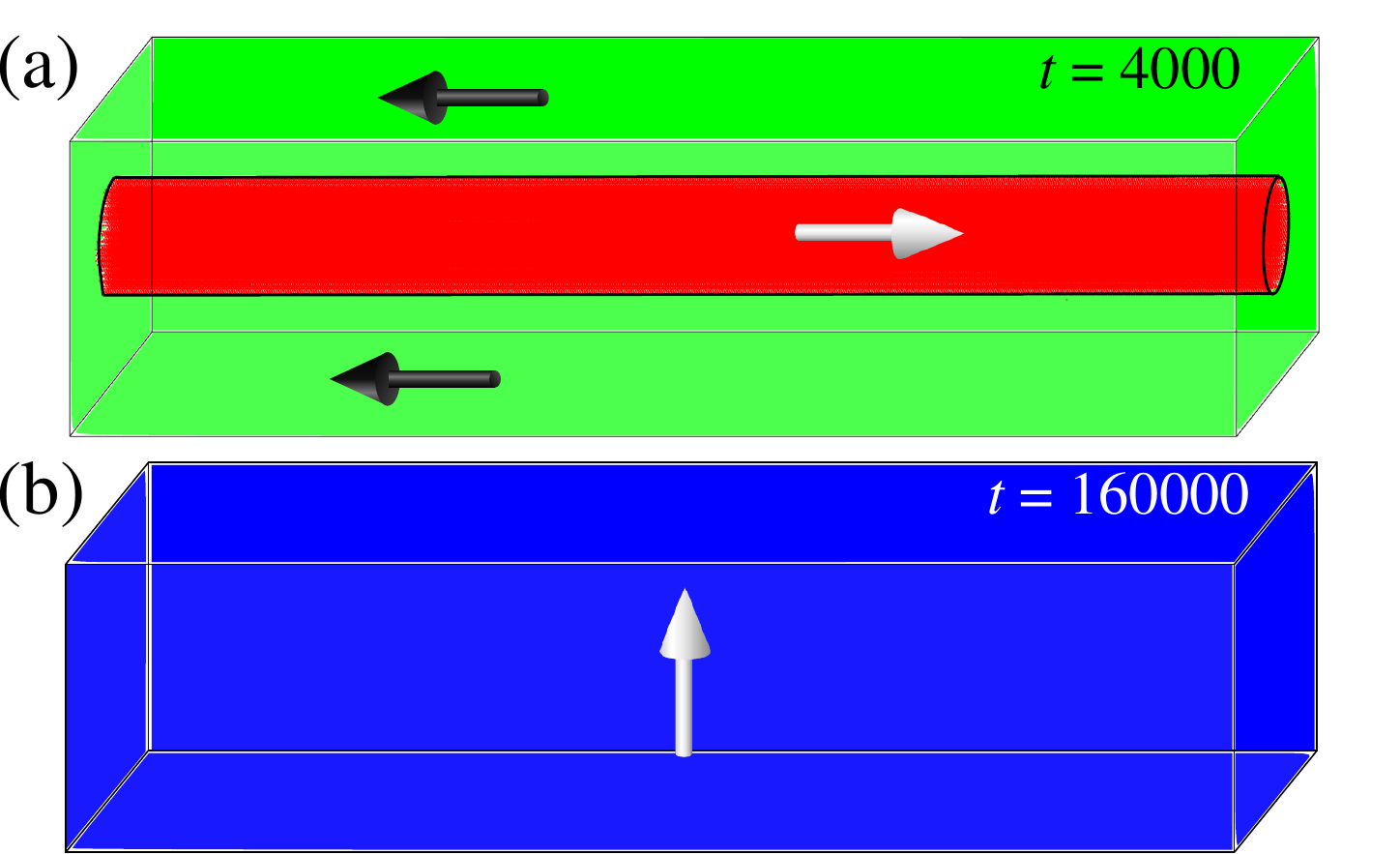}
    \caption{\textbf{Steady states following droplet excitation in the 6-state 3d-APM at high self-propulsion.} (a) \textit{Counterpropagating droplet:} $\sigma =1$ droplet (red) in $\sigma=3$ liquid (green) at $\epsilon=4.5$ results in a high-density droplet state (red) as a cylinder piercing through the original state in the background (green). (b) \textit{Transversely propagating droplet:}  $\sigma =2$ droplet (blue) in $\sigma=3$ (green) liquid at $\epsilon=3.5$ rapidly expands, altering the initial liquid ($\sigma =3$) to $\sigma=2$ liquid phase. Parameters: $\beta = 1$, $\rho_0 = 10$, $r_d = 10$, $\rho_0^d = 1.2\rho_0$, $L_x = 500$, and $L_y = L_z = 50$.}
    \label{fig:3D6smeta}
\end{figure}

\begin{figure}[!t]  
\centering
\includegraphics[width=\columnwidth]{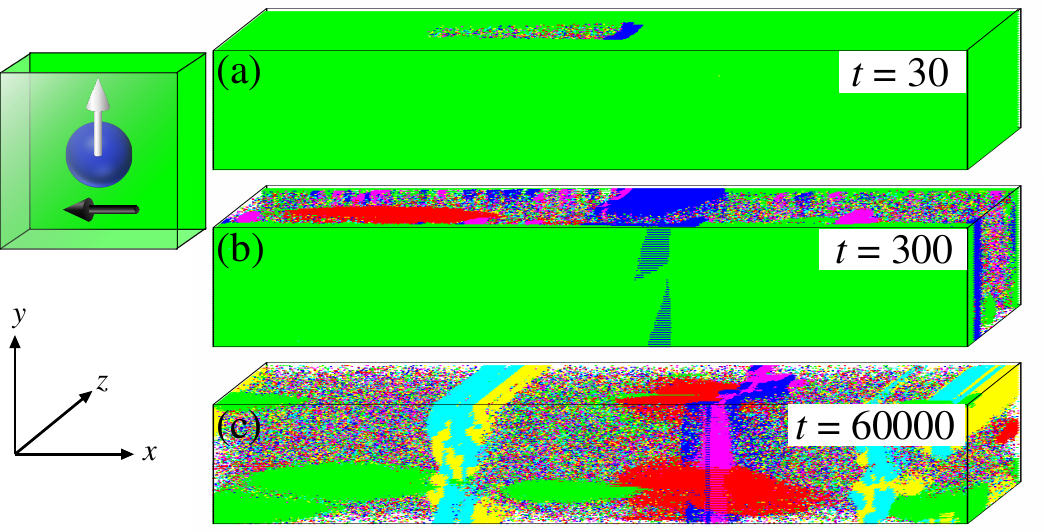}
\caption{\textbf{Time evolution following transverse droplet excitation in the 6-state 3d-APM at very high self-propulsion ($\epsilon=4.5$).} A $\sigma =2$ droplet (blue) introduced in a $\sigma=3$ liquid (green) background. (a) At $t=30$, the initial droplet begins to disrupt the uniform background. (b) At $t=300$, the perturbation stretches rapidly along the $xz$-plane. (c) At $t=60000$, the system shatters into irregular clusters instead of completely altering the initial phase like for lower bias ($\epsilon=3.5$) [Fig.~\ref{fig:3D6smeta}]. Color mapping: right ($\sigma = 1$, red); up ($\sigma = 2$, blue); left ($\sigma = 3$, green); down ($\sigma = 4$, magenta); front ($\sigma = 5$, yellow); behind ($\sigma = 6$, cyan). Parameters: $\beta = 1$, $\rho_0 = 10$, $r_d = 10$, $\rho_0^d = 1.2\rho_0$, $L_x = 500$, and $L_y = L_z = 50$.}
\label{fig:S1}
\end{figure}

Fig.~\ref{fig:3D6smeta} captures the time evolution and steady-state spatial morphologies deep inside the liquid phase ($\beta=1, \rho_0 =10$) for the 6-state 3d-APM at high self-propulsion. When a counter-propagating liquid droplet ($\sigma=1$) is introduced, transverse hopping is suppressed in all directions orthogonal to the propagation axis ($y$ and $z$). Consequently, the droplet evolves into a cylindrical state [Fig.~\ref{fig:3D6smeta}(a)]. This contrasts with the 4-state model, where unsuppressed pure diffusion along the $z$-axis allows the droplet to spread vertically, resulting in a planar \textit{slab} state [Fig.~\ref{fig:3D4smeta}(d)].

Conversely, a transversely propagating droplet ($\sigma=2$) can completely destroy the initial ordered background ($\sigma=3$), culminating in a macroscopic phase reversal to a $\sigma=2$ liquid [Fig.~\ref{fig:3D6smeta}(b)]. However, achieving this full reversal requires a slightly lower bias (e.g., $\epsilon=3.5$) to maintain sufficient transverse hopping for the droplet to expand laterally. 

At extremely high self-propulsion ($\epsilon=4.5$), the non-preferred transverse hopping rate approaches zero. Consequently, a transversely propagating droplet cannot easily expand laterally across the $yz$-plane to conquer the background, as it does during the full phase reversal at a slightly lower bias ($\epsilon=3.5$). Instead, the disruption caused by the droplet is immediately carried by the fast longitudinal flow of the surrounding background particles [Fig.~\ref{fig:S1}(a)]. Because active motion along the $x$-axis is highly biased while transverse hopping is severely suppressed, the perturbation rapidly stretches along the $x$-axis rather than spreading through the $yz$-plane [Fig.~\ref{fig:S1}(b)]. This extreme directional imbalance prevents uniform lateral expansion and permanently shatters the initial uniform flock into a highly fragmented, multi-domain configuration of competing, irregular clusters [Fig.~\ref{fig:S1}(c)].

\bibliographystyle{apsrev4-2}
\bibliography{bibmanuscript_3DAM}
\end{document}